# Optimized Photoemission from Organic Molecules in 2D Layered Halide Perovskites

Muhammad S. Muhammad, Dilruba A. Popy, Hamza Shoukat, John M. Lane, Neeraj Rai, Vojtěch Vaněček, Zdeněk Remeš, Romana Kučerková, Vladimir Babin, Chenjia Mi, Yitong Dong, Mark D. Smith, Novruz G. Akhmedov, Daniel T. Glatzhofer, and Bayram Saparov*

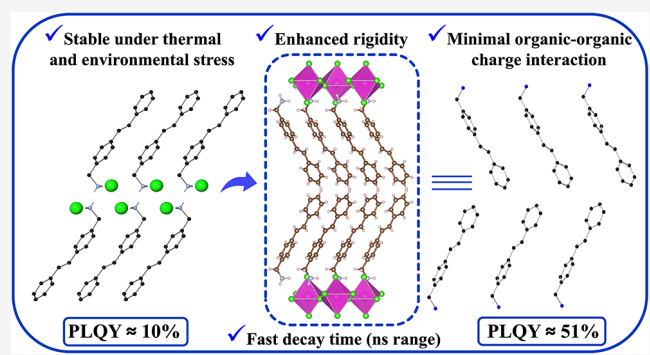

**ABSTRACT:** In recent years, hybrid organic−inorganic metal halides have been at the forefront of materials research. Typically, the functional (e.g., optoelectronic) properties of hybrid halides are derived from the inorganic structural part, whereas the organic structural units can add extra advantages in terms of stability, rigidity, and processability. Here, we report the design, synthesis, and characterization of two new hybrid materials in which the outstanding photophysical properties originate from the organic structural part. The new compounds, $(C_{15}H_{16}N)_2CdCl_4$ and $((Br)C_{15}H_{15}N)_2CdCl_4$, have 2D layered Ruddlesden−Popper-type perovskite structures. These hybrids are blue-white light emitters just like their corresponding pure organic salts, but with much improved emission efficiencies. Optical spectroscopy and density functional theory (DFT) studies confirm that photoemission comes from the *trans*-stilbene organic cations. The photoluminescence quantum yield (PLQY) values of these new materials are among the highest known, 50.83% and 26.60% for $(C_{15}H_{16}N)_2CdCl_4$ and $((Br)C_{15}H_{15}N)_2CdCl_4$, respectively. This is up to a 5-fold increase as compared to the light emission efficiency of the precursor salt $C_{15}H_{16}NCl$ (PLQY of 10.33%). Alongside their outstanding optical properties, their environmental and thermal stability allow their consideration for potential practical applications such as radiation detection. This work shows that hybrid metal halides can be compositionally and structurally engineered to have highly efficient photoemission originating from the organic components for fast scintillation applications.

## 1. INTRODUCTION

In recent years, metal halides have been at the forefront of materials research due to their remarkable crystal chemistry, optical, and electronic properties.[1−9] Owing to their vast compositional phase space, material design and synthesis efforts are seemingly limited only to researchers' imagination and creativity. All-inorganic compositions such as $CsPbX_3$ (X = Cl, Br, I), alkali copper halides, manganese-based compositions, and silver-based halides have been extensively studied, which lead to the discoveries of multiple families of materials with excellent photophysical properties.[1−4,10−15] On the other hand, a wide range of organic−inorganic hybrid materials have also been reported to have exciting properties that can enable their use in applications ranging from photodetection to light-emitting diodes, and solar cell applications.[16−24] Most well-known all-inorganic and hybrid organic−inorganic metal halides owe their functional photophysical properties to the inorganic anionic components that can be zero-dimensional (0D) molecules, one-dimensional (1D) chains, two-dimensional (2D) layers, and three-dimensional (3D) networks.

Taking advantage of chemical composition and structural tunability of metal halides, hybrid organic−inorganic halides with an emissive organic structural part can be designed. Although this line of research remains underexplored, it can lead to preparation of novel optical materials with key advantages not seen in inorganic light emitters. To achieve this, a deliberate attempt must be made to select suitable metal and halogens that will give very large inorganic band gaps. In such cases, the use of low highest occupied molecular orbital (HOMO)—lowest unoccupied molecular orbital (LUMO) gap organic molecules can lead to the preparation of hybrid halides with organic frontier orbitals (Figure S1). As a result, the overall band gap and photophysical properties of these hybrids including their photoluminescence excitation (PLE) and photoluminescence emission (PL) profiles are dictated by the organic component. So far, there are only a few examples of the demonstration of this materials design concept. For example, Fattal et al. reported a hybrid organic−inorganic indium bromide containing trimethyl(4-stilbenyl)methylammonium cation, which demonstrates blue light emission originating from the organic structural part only.[25] The use of electropositive group 12 metals such as Zn and Cd has also been shown to yield hybrid halides with an all-organic emission.[26,27] Another advantage of the use of Zn and Cd halide-based hybrids is their solubility in common solvents such as methanol used for crystal growth. While these literature studies support the validity of the outlined materials design concept, the resultant hybrid materials with the target organic photoemission have lower light emission efficiencies with the reported photoluminescence quantum yield (PLQY) values of only up to 25%. These values are far lower than PLQYs of up to 100% reported for metal halide light emitters in which the inorganic component is photoactive.[23,28−31] Therefore, further composition and structural optimization are necessary to improve the efficiency of organic emissions in hybrid metal halides.

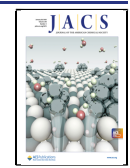

Scheme 1. Retrosynthetic Strategy Used for the Preparation of the Organic Salt $(R)C_{15}H_{15}NCl^i$

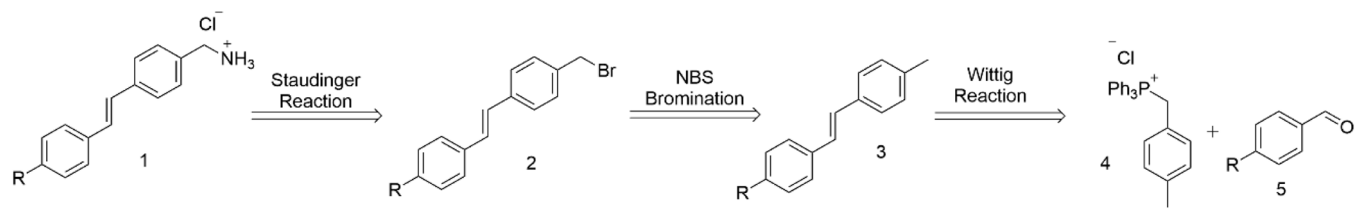

[i]R is either bromine (a) or hydrogen (b).

Hybrid metal halides with exceptional optical properties are currently attracting global attention due to their prospective applications including displays, lighting, and radiation detection. For certain applications, there is a clear advantage for materials exhibiting fast organic emissions as they are candidates for fast scintillation radiation detection.[32] This work reports a significant advancement in optimization of the organic photoemission from hybrid materials. While earlier studies using stilbene-based organics lead to promising hybrid materials with organic photoemissive components, the emission efficiencies remained modest with PLQYs below 20%.[25,26] In these materials, the lower PLQYs are tentatively attributed to aggregation-caused quenching (ACQ).[33] Herein, we further modified the organic molecular components used in earlier studies by removing the cationic nitrogen atom from the ring of (E)-4-styrylpyridinium, which was used in study by Creason and coworkers,[26] and replaced the three (3) methyl groups on the trimethyl(4-stilbenyl)methylammonium, which was the cation of choice for Fattal et al.,[25] with hydrogen to reduce the bulkiness around the cation. The latter makes it possible for the ammonium functional group to reside into pockets formed by the inorganic perovskite layers. The reaction of the resultant (E)-(4-styrylphenyl)methanaminium chloride (($C_{15}H_{16}NCl$) and (E)-(4-(4-bromostyryl)phenyl)methanaminium chloride (($Br)C_{15}H_{15}NCl$) precursors with the large band gap inorganic halide $CdCl_2$ lead to the preparation of two new 2D layered Ruddlesden−Popper-type perovskites $(C_{15}H_{16}N)_2CdCl_4$ and $((Br)C_{15}H_{15}N)_2CdCl_4$, which demonstrate optimized organic emissions. The bluish-white room temperature emission is visibly bright with the measured PLQY values of 50.83% and 26.60% for $(C_{15}H_{16}N)_2CdCl_4$ and $((Br)C_{15}H_{15}N)_2CdCl_4$, respectively. The PLQY of 50.83% is particularly notable as this constitutes a nearly 5-fold increase as compared to the light emission efficiency of the precursor organic salt $C_{15}H_{16}NCl$ (PLQY of 10.33%), and to the best of our knowledge, this is among the highest values of PLQY reported for Cd-based hybrid halides. In fact, this is the highest for Cd-based hybrid halides in which light emission originates from the organic structural part. We discuss the origin of the PLQY improvement in the reported compounds through a combination of X-ray crystallography, optical spectroscopy and Density Functional Theory (DFT) calculations. Our results demonstrate the strong potential of these materials for scintillation applications, particularly given their excellent afterglow characteristics, which surpass those of several state-of-the-art scintillators. Importantly, this work shows that fine-tuning of chemical compositions and crystal structures of hybrid halides via organic molecular design can lead to ultrahigh efficiency emission originating from the organic components. The PLQYs of the resultant materials can rival that of the hybrid metal halides in which emission originates from the inorganic component, but advantageously, the fast photoemission of organic emitters is crucial for fast scintillation applications.

## 2. EXPERIMENTAL SECTION

### 2.1. Materials

Methanol (Sigma-Aldrich), cadmium(II) chloride (79.80%, Fisher Scientific), triphenylphosphine (99%, Aldrich), hydrochloric acid (37%, Sigma-Aldrich), benzaldehyde (>99.0%, Aldrich), para-bromobenzaldehyde (95%, Maybridge), ethanol (>99.99%, Pharmco), sodium hydroxide (95%, EMD), N-bromosuccinimide (99%, Alfa Aesar), benzoyl peroxide (75%, Spectrum), benzene (>99.0%, Sigma-Aldrich), toluene (>99.5%, Sigma-Aldrich), tetrahydrofuran (>99.0%, Sigma-Aldrich), sodium azide (99.5%, Sigma), sodium carbonate (99.5%, Mallinckrodt Chemicals), magnesium sulfate (T. J. Baker), dichloromethane (>99.5%, Sigma-Aldrich), and dimethyl sulfoxide (Fisher Chemicals) were purchased and used with no further purification. Unless otherwise stated, all synthesis experiments were performed in a fume hood under standard conditions. Additional experimental details including syntheses of the organic precursors, Nuclear Magnetic Resonance (NMR) spectra etc. are detailed in the (Supporting Information SI).

### 2.2. Retrosynthesis of $((Br)C_{15}H_{15}N)_2CdCl_4$ and $(C_{15}H_{16}N)_2CdCl_4$

Scheme 1 illustrates the retrosynthetic pathway for the formation of the precursor organic salts. Synthetic procedures are provided in the SI (3a, 2a, and 1a: R = Br and 3b, 2b, and 1b: R = H). The precursor ammonium salts 1 were obtained from Staudinger reactions on benzylic bromides 2. Benzylic bromides 2 were obtained by N-bromosuccinimide (NBS) bromination of methylstilbenes 3. Methylstilbenes 3 were synthesized by Wittig reaction of (4-methylbenzyl)- triphenylphosphonium chloride (4) with the appropriate benzaldehyde 5.

**2.2.1. Formation of ((Br)C$_{15}$H$_{15}$N)$_2$CdCl$_4$.** Ammonium salt 1a ((Br)C$_{15}$H$_{15}$NCl) (0.162 g, 0.560 mmol) was dissolved in 20 mL methanol in a small beaker. In another beaker, 0.057 g (0.250 mmol) of CdCl$_2$·2.5H$_2$O was dissolved in 25 mL of methanol and warmed up to about 50 °C until the formation of a clear solution. The solutions were combined, followed by the addition of two to three drops of 12N hydrochloric acid. After a few minutes, the formation of a white polycrystalline sample of ((Br)C$_{15}$H$_{15}$N)$_2$CdCl$_4$ was observed. The product was collected by gravity filtration and washed with a small amount of methanol. The remaining saturated solution of the hybrid material was slowly evaporated at room temperature to give a second crop of crystals. The combined products (0.143 g, 69%) were stored at room temperature and ambient conditions for further analysis. Crystals of sufficient quality were chosen for SCXRD analysis.

**2.2.2. Formation of (C$_{15}$H$_{16}$N)$_2$CdCl$_4$.** Ammonium salt 1b (C$_{15}$H$_{16}$NCl) (0.100 g, 0.418 mmol) was dissolved in 10 mL methanol in a small beaker. In another beaker, 0.045 g (0.197 mmol) of CdCl$_2$·2.5H$_2$O was dissolved in 20 mL of methanol with warming to form a clear solution. The solutions were combined, followed by the addition of two to three drops of 12N hydrochloric acid. After a few minutes, the formation of a white shiny polycrystalline precipitate of (C$_{15}$H$_{16}$N)$_2$CdCl$_4$ was observed. The product was collected by gravity filtration and washed with excess methanol. The remaining saturated solution of the hybrid material was slowly evaporated at room temperature to collect more crystals. The combined products (0.097 g, 73%) were stored at room temperature and ambient conditions for further analysis. Crystals of sufficient quality were isolated for SCXRD analysis.

### 2.3. Nuclear Magnetic Resonance Measurements

$^1$H NMR spectra were recorded on a 400 MHz VNMRS spectrometer at 25 °C operating at 399.78 MHz equipped with a 5 mm X(H–F) PFG (z-axis pulsed field gradient) Probe. Sample concentration was 10 mg/0.7 mL. Free induction decay (FID) of the samples was processed using the commercially available NMR software package MNOVA 16, https://mestrelab.com/main-product/nmr. The $^1$H NMR chemical shifts are given relative to the relative residual proton peak of the solvents used (7.26 ppm for CDCl$_3$ and 2.5 ppm for DMSO-$d_6$). Typical parameters for acquiring $^1$H NMR spectra were as follows: spectral width 7807.16 Hz, acquisition time 4.0 s, pulse width 3.16 μs (45°), relaxation time 2 s, and number of transients 16.

### 2.4. Preparation of Polymer Films of (C$_{15}$H$_{16}$N)$_2$CdCl$_4$ and ((Br)C$_{15}$H$_{15}$N)$_2$CdCl$_4$

Poly(methyl methacrylate) (PMMA, 1.0 g) was mixed with toluene (3.0 mL) in a 20 mL glass vial and stirred for 7–8 h at room temperature to obtain a homogeneous solution. In parallel, (C$_{15}$H$_{16}$N)$_2$CdCl$_4$ (200 mg) and ((Br)C$_{15}$H$_{15}$N)$_2$CdCl$_4$ (200 mg) were ground into fine powder samples. Each sample of ((C$_{15}$H$_{16}$N)$_2$CdCl$_4$ and ((Br)C$_{15}$H$_{15}$N)$_2$CdCl$_4$) was introduced separately into 1.5 mL of PMMA/toluene solution, and the mixtures were stirred overnight to ensure complete dispersion. The resulting inks were used to fabricate polymer composites. The composites were deposited on cleaned microscopic glass slides (22 mm) by spin-coating 200–300 μL of the subject solution (200 rpm for 30 s) using a Laurell WS-650MZ-23NPPB spin coater. The composites were dried overnight under ambient conditions before characterization.

### 2.5. Powder X-ray Diffraction (PXRD) Measurements

Powder X-ray diffraction (PXRD) measurements were carried out at ambient temperature using a Rigaku MiniFlex600 system with a Ni-filtered Cu Kα radiation source. PXRD scans were performed on polycrystalline samples in the 2–90° (2θ) range with a 0.02° step size. The XRD patterns were analyzed using a PDXL2 software package. The obtained PXRD patterns were fitted using the decomposition method.

### 2.6. Single-Crystal X-ray Diffraction (SCXRD) Measurements

X-ray intensity data from colorless rectangular plates were collected at 100(2) K using a Bruker D8 QUEST diffractometer equipped with a PHOTON-II area detector and an Incoatec microfocus source (Mo Kα radiation, λ = 0.71073 Å). The crystal-to-detector distance was set at 110 mm for (C$_{15}$H$_{16}$N)$_2$CdCl$_4$ and 70 mm for ((Br)-C$_{15}$H$_{15}$N)$_2$CdCl$_4$ to increase the observed diffraction spot separation. The raw area detector data frames were reduced, scaled, and corrected for absorption effects using the Bruker APEX3, SAINT+, and SADABS programs.[34,35] The structure was solved with SHELXT.[36] Subsequent difference Fourier calculations and full-matrix least-squares refinement against $F^2$ were performed with SHELXL-2019/3[36] using OLEX2.[37] Details of the data collection and crystallographic parameters are given in Table S1. Atomic coordinates, equivalent isotropic displacement parameters, selected interatomic distances, and bond angles are provided in Tables S2–S5. The Crystallographic Information Files (CIFs) were deposited in the Cambridge Crystallographic Data Centre (CCDC) database (deposition numbers 2463650 and 2463654–2463655).

### 2.7. Thermogravimetric Analysis and Differential Scanning Calorimetry

Thermogravimetric analysis and differential scanning calorimetry (TGA/DSC) measurements were performed on 5–8 mg samples of the new hybrid compounds and the respective precursor organic salts used in their preparation on a TA Instruments SDT 650 thermal analyzer system. Crystals were heated from 25 to 475 °C under an inert atmosphere of nitrogen gas flow at a rate of 100 mL/min and a heating rate of 5 °C/min. Melting point measurements for both salts (C$_{15}$H$_{16}$NCl and (Br)C$_{15}$H$_{15}$NCl) and the hybrids ((C$_{15}$H$_{16}$N)$_2$CdCl$_4$ and ((Br)C$_{15}$H$_{15}$N)$_2$CdCl$_4$) were taken on a Mel-Temp apparatus (110/120VAC; 50/60 Hz and 200 W). The heating element was initially set at 50 V and later increased to 60 V. Measurements took approximately 35 min for the salts and 40 min for the hybrids, using capillary tubes (0.8–1.1 × 90 mm).

### 2.8. Photoluminescence, Radioluminescence, and Afterglow Measurements

Room-temperature photoluminescence emission (PL) and photoluminescence excitation (PLE) measurements were carried out on polycrystalline samples of the organic salts and hybrid compounds using HORIBA Jobin Yvon Fluorolog-3 spectrofluorometer with xenon lamp source and Quanta-φ integrating sphere. Data were collected using the two-curve method ranging from 250 to 750 nm. For the photostability measurement, samples were placed inside the Quanta-φ integrating sphere on the Jobin Yvon Fluorolog-3 spectrofluorometer. The samples were then exposed to the full power of the xenon lamp at the respective photoluminescence excitation maximum wavelengths of the compounds. Periodic PLQY measurements were taken every 5 min under these conditions for 60 min.

The PL lifetime was measured using a time-correlated single-photon counting method on two independent setups:

> A customized epi-illuminating fluorescence microscope. The sample was excited with a 405 nm pulsed laser (PicoQuant LDH D-C-405) driven by a PicoQuant Sepia PDL-828 driver, operating at a 5 MHz repetition rate. The fluorescence was collected by the objective (Olympus UPLXAPO100XO, NA = 1.45), passed through a series of optical filters to remove the residue laser, and sent to a single-photon avalanche photodiode (Hamamatsu SPAD module C11202–100) connected to a time correlator (PicoQuant HydraHarp 400). The instrument response function is ~250 ps.

> A custom-made spectrofluorometer 5000 M (Horiba Jobin Yvon, Wildwood, MA, USA) equipped with a nanosecond nanoLED pulsed excitation light source. The detection part of the setup involved a single-grating monochromator and a photon-counting detector TBX-04 (IBH Scotland). The convolution procedure was applied to the photoluminescence decay curves to determine true decay times (SpectraSolve software package, Ames Photonics).

Radioluminescence (RL), afterglow, photoluminescence excitation (PLE), and photoluminescence emission (PL) spectra were measured at RT by a custom-made spectrofluorometer 5000 M (Horiba Jobin Yvon, Wildwood, MA, USA) using the tungsten X-ray tube (40 kV, 15 mA, Seifert) and steady-state xenon lamp (EQ-99X LDLS—Energetiq, a Hamamatsu Company) as the excitation sources. The detection part of the setup consisted of a single-grating monochromator and a photon-counting detector TBX-04 (IBH Scotland). Measured spectra were corrected for the spectral dependence of detection sensitivity (RL, PL) and excitation light spectral dependence (PLE). The RL spectrum of BGO ($Bi_4Ge_3O_{12}$) reference powder scintillator was measured under identical geometrical conditions to obtain quantitative information on RL intensity of the samples.

### 2.9. Diffuse Reflectance Measurements

UV−vis diffuse reflectance data were collected on gently ground powder samples using a PerkinElmer LAMBDA 750 UV−vis-NIR spectrometer with a 100 mm InGaAs Integrating Sphere over 250−1100 nm range. Diffuse reflectance data were then transformed to pseudoabsorption spectra employing the Kubelka−Munk function $F(R) = \alpha/S = (1-R)^2/2R$, where $\alpha$ refers to the absorption coefficient, $S$ is the scattering coefficient, and $R$ is the reflectance.

### 2.10. Photothermal Deflection Spectroscopy

Absorbance spectra were also measured in the 300−1400 nm spectral range using a photothermal deflection spectroscopy (PDS) setup.[38,39] PDS measures optical absorption indirectly by detecting the deflection of a probe laser beam caused by thermally induced refractive index changes in the medium near the sample surface. It is an extremely sensitive, noncontact method for detecting weak absorption. During the optical measurements, the samples were immersed in liquid (Florinert FC72, 3 M Company, St. Paul, MN, USA) to measure the optical absorption independently for selected photon energies. Quazi monochromatic light was provided by a 150 W Xe lamp and a monochromator (SpectraPro-150, Acton Research Corp., Acton, MA, USA) equipped with two gratings: a UV holographic grating (1200/mm) and a ruled grating (600/mm) blazed at 500 nm and with slits of 1/1 mm. The spectral resolution was 5 nm for the UV holographic grating and 10 nm for the ruled grating. The spectra were calibrated by measuring the PDS of a black carbon sample.

### 2.11. Computational Studies

All theoretical calculations were performed using the Vienna Ab-initio Simulation Package[40−44] with the projector-augmented-wave method. For relaxation, the Perdew−Burke−Ernzerhof (PBE) functional was used with Grimme's D3 corrections.[45,46] Both $(C_{15}H_{16}N)_2CdCl_4$ and $((Br)C_{15}H_{15}N)_2CdCl_4$ were relaxed in the unit cell before other calculations were performed. A primitive cell was found for $((Br)C_{15}H_{15}N)_2CdCl_4$ and used in all reported calculations. Using Monkhorst−Pack,[47] a k-point mesh of 3 × 3 × 1 was generated for $(C_{15}H_{16}N)_2CdCl_4$ and a mesh of 3 × 3 × 3 was used for $((Br)C_{15}H_{15}N)_2CdCl_4$. For the sake of increased accuracy, band structure calculations were carried out using the Heyd−Scuseria−Ernzerhof 2006 (HSE06) functional,[48] which employs 25% Hartree−Fock exact exchange. 500 eV was used as the plane wave basis kinetic energy cutoff, with convergence criteria for the self-consistent field (SCF) set to $10^{-5}$ eV for the energies and 0.02 eV Å$^{-1}$ for the residual forces.

### 2.12. Scintillation Decay Kinetics

For measurement of the fast scintillation decays with high resolution, the samples were excited by picosecond (ps) X-ray tube N5084 from Hamamatsu, operating at 40 kV. The X-ray tube is driven by the ps light pulser equipped with a laser diode operating at 405 nm. The repetition rate can go up to 10 MHz. The adjustable delay generator is triggering the laser pulses and the detector readout. The signal was detected by a hybrid picosecond photon detector HPPD-860 and Fluorohub unit (time-correlated single-photon counting technique) from Horiba Scientific. The instrumental response function full width at half maximum (fwhm) of the setup is about 76 ps. Samples are mounted a few centimeters in front of the beryllium window of the X-ray tube with a 45° angle to the incident beam. The luminescence is detected from the same surface by the detector. The convolution procedure was applied to the photoluminescence decay curves to determine true decay times (SpectraSolve software package, Ames Photonics).

## 3. RESULTS AND DISCUSSION

### 3.1. Crystal Structures

The fine plate-like crystals of both $(C_{15}H_{16}N)_2CdCl_4$ and $((Br)C_{15}H_{15}N)_2CdCl_4$ can be synthesized by mixing the methanol solutions of the corresponding organic salts and $CdCl_2$. A schematic representation of the synthesis method is shown in Scheme 2, and a detailed explanation of the synthesis procedures is given in Section 2.2. The phase purity and crystallinity of the as-synthesized samples of both hybrid compounds have been confirmed by comparing the powder and single-crystal X-ray diffraction data (Figure S2). $(C_{15}H_{16}N)_2CdCl_4$ and $((Br)C_{15}H_{15}N)_2CdCl_4$ are formed as shiny white crystals under daylight and emit bright blue light under UV irradiation. The blue light emitted by $(C_{15}H_{16}N)_2CdCl_4$ is more pronounced to the naked eye compared to that emitted by $((Br)C_{15}H_{15}N)_2CdCl_4$.

**Scheme 2. Schematic Representation for the Synthesis Process of $(C_{15}H_{16}N)_2CdCl_4$ and $((Br)C_{15}H_{15}N)_2CdCl_4$**

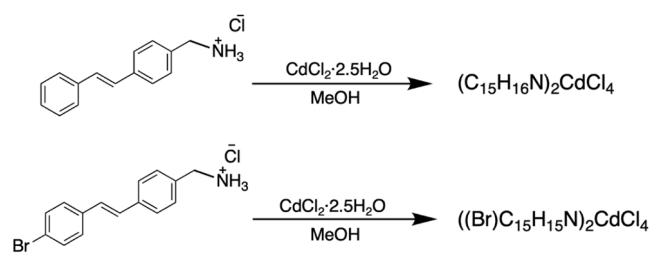

Single-crystal X-ray diffraction (SCXRD) experiments suggest that $(C_{15}H_{16}N)_2CdCl_4$ and $((Br)C_{15}H_{16}N)_2CdCl_4$ have Ruddlesden−Popper-type layered crystal structures. Both compounds crystallized in the orthorhombic crystal system with a space group of $Pbca$ for $(C_{15}H_{16}N)_2CdCl_4$ and a space group of $Aea2$ for $((Br)C_{15}H_{15}N)_2CdCl_4$ (Table S1). The cadmium chloride $CdCl_6$ octahedra grow in a two-dimensional fashion connected to each other in two directions through corner sharing (see Figures 1 and 2) to form the polyanionic $[CdCl_4]^{2-}_\infty$ layers. The organic cations regularly stack themselves owing to the flat nature of all their sp$^2$ hybridized carbons, except for the carbon attached to the amine group. The unit cells of both crystal structures are elongated in the interlayer directions, which is due to the presence of the large organic cations. One would expect to have a slightly larger unit cell in $((Br)C_{15}H_{15}N)_2CdCl_4$ because of the addition of bromine in the para position of the organic molecule. However, the packing of the (Br)-$C_{15}H_{15}N^+$ molecules are offset along the $a$-direction in $((Br)C_{15}H_{15}N)_2CdCl_4$, which together with the halogen-bonding interactions between Br atoms ensure a slightly smaller interlayer lattice parameter in $((Br)C_{15}H_{15}N)_2CdCl_4$ (Table S1 and Figure S3). Within the organic cationic bilayer, the distance between the layers in $(C_{15}H_{16}N)_2CdCl_4$ is found to be 4.12 Å (Figure S4). In comparison, the corresponding distance in $((Br)C_{15}H_{15}N)_2CdCl_4$ is found to be 3.82 Å. This reduced distance is within the range of 3.4 to 3.9 Å for a halogen−halogen interaction reported in literature.[49]

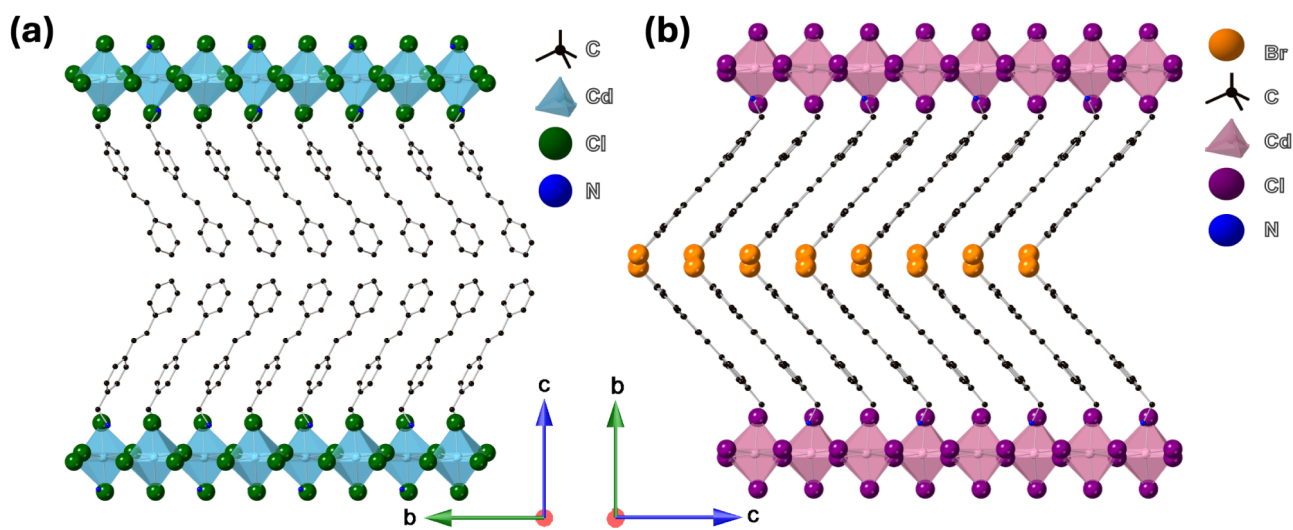

**Figure 1.** Polyhedral representations of the crystal structures along *a*-axis for (a) $(C_{15}H_{16}N)_2CdCl_4$ and (b) $((Br)C_{15}H_{16}N)_2CdCl_4$.

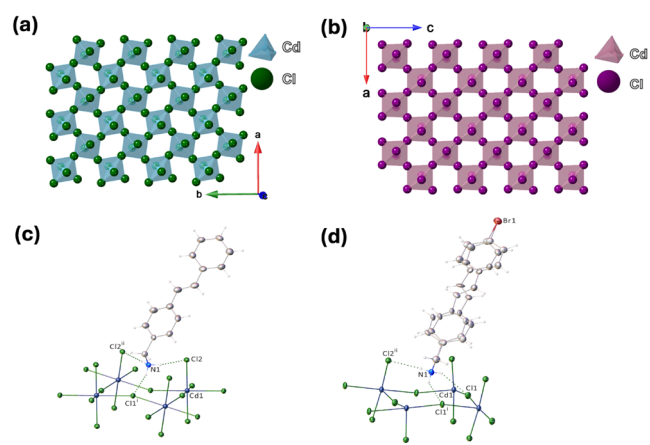

**Figure 2.** Polyhedral representations of 2D inorganic layers in (a) $(C_{15}H_{16}N)_2CdCl_4$ and (b) $((Br)C_{15}H_{16}N)_2CdCl_4$. Hydrogen bonding contacts in (c) $(C_{15}H_{16}N)_2CdCl_4$ and (d) $((Br)C_{15}H_{16}N)_2CdCl_4$. "Stilbene" part of the cation disordered over two orientations, major component occupancy fraction $A = 0.56(1)$. Displacement ellipsoids are drawn at the 50% probability level for (c) and (d).

For $(C_{15}H_{16}N)_2CdCl_4$, the 2D $[CdCl_4]_\infty^{2-}$ layer extends along the *ab*-plane, while stilbene cations are arranged vertically along the *c*-axis, stabilizing the charges of the inorganic anionic sheets. Within the inorganic layers, $[CdCl_6]^{4-}$ octahedra are slightly distorted with the Cd−Cl bond lengths ranging from 2.5722(11) Å to 2.6533(11) Å (Table S5). The $MX_n$ polyhedral distortion in metal halides has been reported for many materials systems.[50,9] For hybrid halides in which the valence (VB) and conduction (CB) bands derive from the inorganic structural component, such distortions can impact the band structures, light absorption, and emission properties.[51−53] However, since the target compounds in this work are designed to have photoactive organic components, the observed inorganic structural distortion is not expected to affect the photophysical properties of $(C_{15}H_{16}N)_2CdCl_4$.

The two benzene rings that are connected by an ethylene group in stilbene have a unique twist conformation (Figure 1) with an angle of 21.89(1)° between the phenyl rings. The observed twist conformation in $(C_{15}H_{16}N)_2CdCl_4$ (Figure S3a) is attributed to stability reasons. The stilbene-based organics are known to either photoisomerize or photodimerize, which leads to PL quenching, and therefore, is the reason for their instability under continuous irradiation.[25,26] However, the distance of 4.12 Å between the organic cationic layers (Figure S4) in the cationic bilayer and the distance of 7.42 Å between the ethylene groups within a single organic layer of $(C_{15}H_{16}N)_2CdCl_4$ are sufficiently long to prevent any light-induced detrimental changes. To understand how the organic units stack themselves without inorganic units, we also performed X-ray diffraction measurements on $C_{15}H_{16}NCl$. The results showed that these organic units packed themselves in a similar way to their arrangements in the corresponding hybrid compound (Figure S5). The distance between the adjacent aromatic rings in $(C_{15}H_{16}N)_2CdCl_4$ is 7.12 Å (Figure S4), while this distance is around 5.13 Å in the precursor organic salt (Figure S5). The increased distance between the organic molecules in the hybrid is very important as it has significant consequences for its photostability and light emission efficiency.

In hybrid materials with the organic photoactive components, subtle changes to the organic cations can have important influences on the photophysical properties. For instance, a recent study showed that the incorporation of Br into the organic cationic layers improves the X-ray detection properties of a new 2D perovskite (R−MPA)(BrEA)PbBr$_4$.[54] With this in mind, we further modified the (*E*)-(4-styrylphenyl)-methanaminium cation used in $(C_{15}H_{16}N)_2CdCl_4$ by incorporating bromine on the para position. The bromine atom is strategically added in that position to maintain the shape of the organic unit, which in turn will ensure proper stacking of the organics needed for the formation of the 2D layered perovskite structure. The resultant hybrid $((Br)C_{15}H_{15}N)_2CdCl_4$ exhibits a layered perovskite structure with the 2D $[CdCl_4]^{2-}$ anionic sheets extended along the *ac*-axes. The $[CdCl_6]^{4-}$ octahedra are more distorted in this case, with the Cd−Cl bond lengths ranging from 2.520(4) Å to 2.728(6) Å. On the other hand, in-plane tilting of $[CdCl_6]^{4-}$ octahedra is less noticeable in $((Br)C_{15}H_{15}N)_2CdCl_4$ (Figure 2a and b). The $(Br)C_{15}H_{15}N^+$ is disordered over two orientations (see SI for further discussions). The presence of a heavy atom, bromine, locks the organic units into place as the bromine atoms alternate, thereby restricting the movement of the benzene ring from assuming the conformation observed in $(C_{15}H_{16}N)_2CdCl_4$ (Figures 1b and S3b). The bromine and $-CH_2NH_3^+$ substituents of the cation are common to both components and do not show disorder in $((Br)C_{15}H_{15}N)_2CdCl_4$. The major disorder component occupancy was refined to 0.56(1). The distance between the closest bromine atoms is 3.827 Å. This distance is short enough for an interaction between the two bromine atoms,[55] which in turn can impact the observed optical properties (discussed later). The 2D layered perovskite structures of $(C_{15}H_{16}N)_2CdCl_4$ and $((Br)C_{15}H_{15}N)_2CdCl_4$ are stabilized through hydrogen bonding interactions between the organic cations and inorganic perovskite sheets (Figure 2c-d). N(H)-Cl hydrogen bond distances were in the range $d(N-Cl)$ = 3.193(5)−3.245(4) Å in $(C_{15}H_{16}N)_2CdCl_4$ and $d(N-Cl)$ = 3.214(7)−3.395(7) Å in $((Br)C_{15}H_{15}N)_2CdCl_4$ (Figure 2). Such hydrogen bonding interactions in layered perovskites are known to increase stability and provide structural rigidity.[9,56]

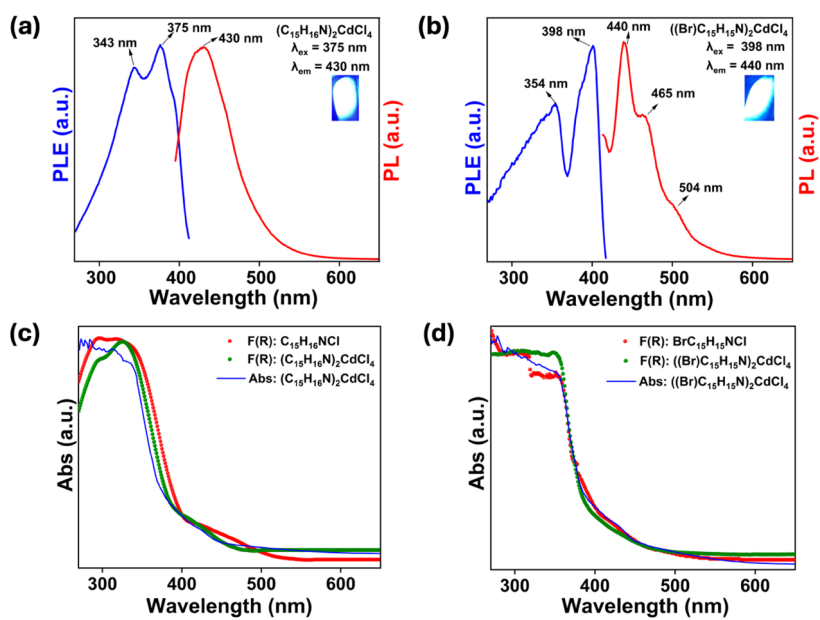

**Figure 3.** PL and PLE spectra and absorption spectra for (a) $(C_{15}H_{16}N)_2CdCl_4$ and (b) $((Br)C_{15}H_{15}N)_2CdCl_4$. Optical absorption spectra obtained from photothermal deflection spectroscopy (Abs, in blue) and diffuse reflectance spectroscopy (F(R), in green) for (c) $(C_{15}H_{16}N)_2CdCl_4$ and (d) $((Br)C_{15}H_{15}N)_2CdCl_4$. The absorption spectra for the corresponding precursor organic salts are also provided for comparison (F(R), in red).

### 3.2. Optical Properties

The photoluminescence excitation (PLE) and emission (PL) as well as absorption spectra of $(C_{15}H_{16}N)_2CdCl_4$ and $((Br)C_{15}H_{15}N)_2CdCl_4$ are provided in Figure 3. The optical spectra obtained for the new hybrid materials confirm the validity of our halide materials with photoactive organics design concept. The compounds demonstrate weak optical absorption below 500 nm, with a much stronger onset of optical absorption below 400 nm. The obtained absorption spectra are similar to that of the corresponding precursor organic salts (Figures 3c-d), suggesting that the organic structural component determines the optical band gaps of the hybrids. This is further confirmed by our computational work (see Figure S6 and accompanying discussion in SI), which suggests that the frontier orbitals of $(C_{15}H_{16}N)_2CdCl_4$ and $((Br)C_{15}H_{15}N)_2CdCl_4$ belong to their respective organic molecules. The two-peak PLE spectra have maxima at 375 and 398 nm for $(C_{15}H_{16}N)_2CdCl_4$ and $((Br)C_{15}H_{15}N)_2CdCl_4$, respectively, leading to blue emissions. The observed multi-band PL emissions range from 350 to 550 nm for $(C_{15}H_{16}N)_2CdCl_4$ and 375 to 600 nm for $((Br)C_{15}H_{15}N)_2CdCl_4$, depending on the excitation wavelengths (Figures 3a-b and S7). To compare, the PL and PLE spectra of the stand-alone organic salts were also measured (Figure S8), revealing bluish-white light emission from the precursor organic salts. Both organic salts show characteristic two PLE peaks around ∼350 and 400 nm, leading to multiband emissions in the 375−600 nm range.

In the case of $C_{15}H_{16}NCl$, the broadband PL emission spectrum has peaks at 405, 429, and 511 nm when excited at 345 nm (Figure S8a). The PL peaks at 405 and 429 nm are typical fluorescence emissions of stilbene and its derivatives.[57−60] The lower energy emission band at 511 nm is attributed to phosphorescence.[59] It has been shown that trans-stilbene itself does not show phosphorescence (Figure S9) but heavy-atom effects due to the presence of halides in its vicinity (e.g., in solution) induce phosphorescence in this spectral region.[58] In the present case, the crystal structures of the organic salts show the close packing of the organic cationic molecules and chloride counterions (Figures S5). Based on these discussions, $(Br)C_{15}H_{15}NCl$ is expected to show a more intense phosphorescence. This salt is heavier due to the presence of bromine at the para position of the inner benzene ring. Indeed, the phosphorescence emission peak is more prominent in the PL spectrum of $(Br)C_{15}H_{15}NCl$ (Figure S8b).[58,59] With the addition of bromine, the PL emission peaks are more red-shifted in $(Br)C_{15}H_{15}NCl$ with a fluorescence emission peak at 444 nm and a phosphorescence emission peak at 539 nm (Figure S8b).

Prior computational studies of stilbenes and their derivatives have shown that their excitation involves transition of an electron from the $\pi-\pi$ antibonding HOMO to the $\pi^*-\pi^*$ bonding LUMO.[61] The absorption bands are typically above 3 eV (<400 nm) but their specific spectral locations are strongly dependent on the substituents on stilbenes and their intermolecular arrangements (Figures S7−S9).[57−61] Even more so, the theoretical studies show that the light emission properties of stilbenes are very sensitive to the molecular packing.[61] Confirming these earlier conclusions, we observe noticeable changes not only between the optical spectra of $C_{15}H_{16}NCl$ and $(Br)C_{15}H_{15}NCl$, but also these organic salts and their corresponding hybrid metal halides. The phosphorescence emission above 500 nm is suppressed in the hybrid materials. This is attributed to the anchoring of the organic cations by the heavy extended inorganic 2D metal halide sheets, which provides a rigid structural framework leading to reduced probability of intersystem crossing.[62,63]

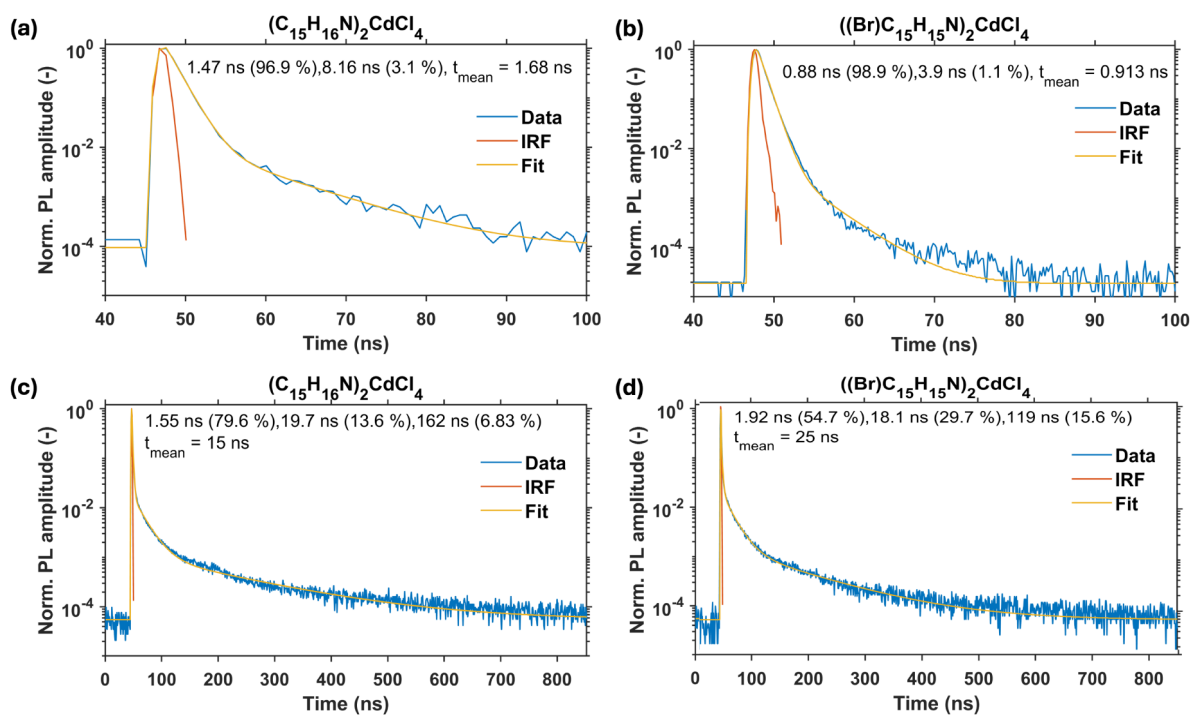

**Figure 4.** Room temperature time-resolved PL for (a, c) $((Br)C_{15}H_{15}N)_2CdCl_4$ and (b, d) $(C_{15}H_{16}N)_2CdCl_4$ collected using an excitation wavelength of 400 nm in short (top) and long (bottom) time windows. The values in parentheses represent intensity-weighted contributions.

The photoluminescence (PL) decay kinetics in Figure 4 exhibit a complex temporal behavior for both $(C_{15}H_{16}N)_2CdCl_4$ and $((Br)C_{15}H_{15}N)_2CdCl_4$ compounds. A biexponential model was required in the short time window to achieve a satisfactory fit. Both compounds display a dominant fast decay component (0.88 ns, 98.9% for $((Br)C_{15}H_{15}N)_2CdCl_4$; 1.47 ns, 96.9% for $(C_{15}H_{16}N)_2CdCl_4$) accompanied by a minor slower component (3.9 ns for $((Br)C_{15}H_{15}N)_2CdCl_4$; 8.16 ns for $(C_{15}H_{16}N)_2CdCl_4$). In the long-time window, a triexponential model was necessary to describe the decay kinetics adequately. In addition to the fastest component, which could not be reliably fitted due to the coarse temporal resolution imposed by the larger bin size (0.832 ns/channel), an intermediate component (18.1 ns for $((Br)C_{15}H_{15}N)_2CdCl_4$; 19.7 ns for $(C_{15}H_{16}N)_2CdCl_4$) and a slow component (119 ns for $((Br)C_{15}H_{15}N)_2CdCl_4$; 162 ns for $(C_{15}H_{16}N)_2CdCl_4$) were identified. Notably, a triexponential PL decay kinetics with similar decay times were reported for stilbene single crystals.[64] The intensity-weighted contributions of each component $c_i$ and mean decay times $\tau_m$ were calculated from the fitting parameters according to the formula:

$$c_i = \frac{A_i \tau_i}{\sum_{i=1}^{n} A_i \tau_i} \quad (1)$$

$$\tau_m = \sum_{i=1}^{n} c_i \tau_i \quad (2)$$

where $\tau_i$ and $A_i$ are the $i^{th}$ decay time and pre-exponential factor, respectively. The mean decay times are 15 and 25 ns for $(C_{15}H_{16}N)_2CdCl_4$ and $((Br)C_{15}H_{15}N)_2CdCl_4$, respectively. The higher mean decay time of $((Br)C_{15}H_{15}N)_2CdCl_4$ is primarily due to a higher contribution of the slow component ($\sim 10^2$ ns), which could be explained by the heavy atom effect (Br substitution into the organic molecule) that promotes the triplet state emission.[58]

The obtained lifetime data in the short time window (Figure 4a-b) are in excellent agreement with the average lifetimes of $\tau_{avg}$ = 1.70 ns and $\tau_{avg}$ = 1.05 ns for $(C_{15}H_{16}N)_2CdCl_4$ and $((Br)C_{15}H_{15}N)_2CdCl_4$, respectively, obtained using a customized epi-illuminating fluorescence microscopy (Figure S10). These ~1 ns fluorescent lifetimes are comparable to that of the corresponding organic salts $C_{15}H_{16}NCl$ and $(Br)C_{15}H_{15}NCl$. The dominant fast fluorescent emission of $(C_{15}H_{16}N)_2CdCl_4$ and $((Br)C_{15}H_{15}N)_2CdCl_4$ could be important for their prospective use in fast radiation detection.

### 3.3. Light Emission Efficiency and Stability

The precursor organic salts used in this study are bluish-white light emitters with measured photoluminescence quantum yield (PLQY) values of 10.33% and 10.46% (Figure S11) for $C_{15}H_{16}NCl$ and $(Br)C_{15}H_{15}NCl$, respectively. These values are higher than the PLQYs of precursor stilbene-based organic salts used in other studies,[25,26] suggesting that our targeted modifications on the organic molecules yielded organic salts that are good emitters on their own. Incorporation of $C_{15}H_{16}N^+$ and $(Br)C_{15}H_{15}N^+$ into a layered perovskite structure significantly increases their light emission efficiencies—PLQYs of 50.83% and 26.60% (Figure S11) for (C$_{15}$H$_{16}$N)$_2$CdCl$_4$ and ((Br)C$_{15}$H$_{15}$N)$_2$CdCl$_4$, respectively, among the highest in this materials class (Table 1). The only higher PLQY value of 52.3% is reported for DMP-1-CdBr$_3$, and the high-efficiency light emission is attributed to self-trapped excitons (STEs) localized on the inorganic structural part.[75] In the case of (C$_{15}$H$_{16}$N)$_2$CdCl$_4$ and ((Br)C$_{15}$H$_{15}$N)$_2$CdCl$_4$, the higher intermolecular separations of organic units compared to the corresponding organic precursor salts result in considerable enhancement of PLQY values and stable and record high emission efficiency organic photoemission in the hybrid materials. The molecular stacking in the stand-alone organic salts C$_{15}$H$_{16}$NCl with distances in the 4.90 Å to 5.14 Å range (Figure S5) is much closer compared to the hybrid material (C$_{15}$H$_{16}$N)$_2$CdCl$_4$ whose analogous distances are between 7.12 Å and 7.42 Å (Figure S4). It is well-known that the vast majority of conjugated planar aromatic organics suffer from excited-state quenching due to ACQ.[76] The significantly increased intermolecular distance in (C$_{15}$H$_{16}$N)$_2$CdCl$_4$ leads to one of the highest reported PLQY for Cd-based hybrid materials (Table 1). The PLQY of 50.83% for (C$_{15}$H$_{16}$N)$_2$CdCl$_4$ is nearly five times higher than the PLQY of 10.33% measured for C$_{15}$H$_{16}$NCl.

Table 1. Photophysical Properties of Some Reported Cadmium-Based Hybrid Organic−Inorganic Halides[26,56,65−74]

| S/N | Compound | PLQY (%) | $PL_{max}$ (nm) | $PLE_{max}$ (nm) | Stokes Shift | Ref |
|---|---|---|---|---|---|---|
| 1 | DMP-1-CdBr$_3$ | 52.3 | 432 | 329 | 103 | 75 |
| 2 | (C$_{15}$H$_{16}$N)$_2$CdCl$_4$ | 50.83 | 430 | 375 | 55 | This work |
| 3 | ((Br)C$_{15}$H$_{15}$N)$_2$CdCl$_4$ | 26.60 | 437 | 354 | 83 | This work |
| 4 | (BAPPz)Cd$_2$Br$_8$·2H$_2$O | 13.78 | 527 | 356 | 171 | 65 |
| 5 | (C$_6$H$_7$ClN)CdCl$_3$ | 12.30 | 530 | 345 | 185 | 66 |
| 6 | [BHEPZ]CdBr$_4$ | 12.00 | 463 | 359 | 104 | 67 |
| 7 | (C$_6$H$_7$NCl)$_2$CdCl$_4$ | 11.70 | 558 | 380 | 178 | 68 |
| 8 | R$_2$CdCl$_4$ | 11.21 | 528 | 453 | 75 | 26 |
| 9 | (H$_3$AEP)$_2$CdBr$_6$·2Br | 9.00 | 612 | 365 | 247 | 74 |
| 10 | [(2-mb)tpp]$_2$CdCl$_4$ | 6.96 | 376 | 335 | 41 | 72 |
| 11 | (C$_6$H$_7$NBr)$_2$CdBr$_4$ | 4.15 | 570 | 380 | 190 | 68 |
| 12 | (P-xd)CdCl$_4$ | 3.97 | 435 | 287 | 148 | 56 |
| 13 | [EPIPZ]CdBr$_4$ | 3.14 | 456 | 368 | 88 | 67 |
| 14 | (2CePiH)CdCl$_3$ | 1.88 | 451(550) | 330 | 121(220) | 69 |
| 15 | (HMEDA)CdCl$_4$ | 1 | 515 | 288 | 227 | 73 |
| 16 | (HMEDA)CdBr$_4$ | 1 | 445 | 365 | 80 | 73 |
| 17 | (R)$_2$CdBr$_4$·DMSO | 0.32 | 501 | 399 | 102 | 71 |
| 18 | (R)CdI$_3$·DMSO | 0.27 | 445 | 515 | 70 | 71 |
| 19 | (C$_6$H$_{14}$N$_2$)CdCl$_4$·H$_2$O | N/A | 443 | 363 | 80 | 70 |

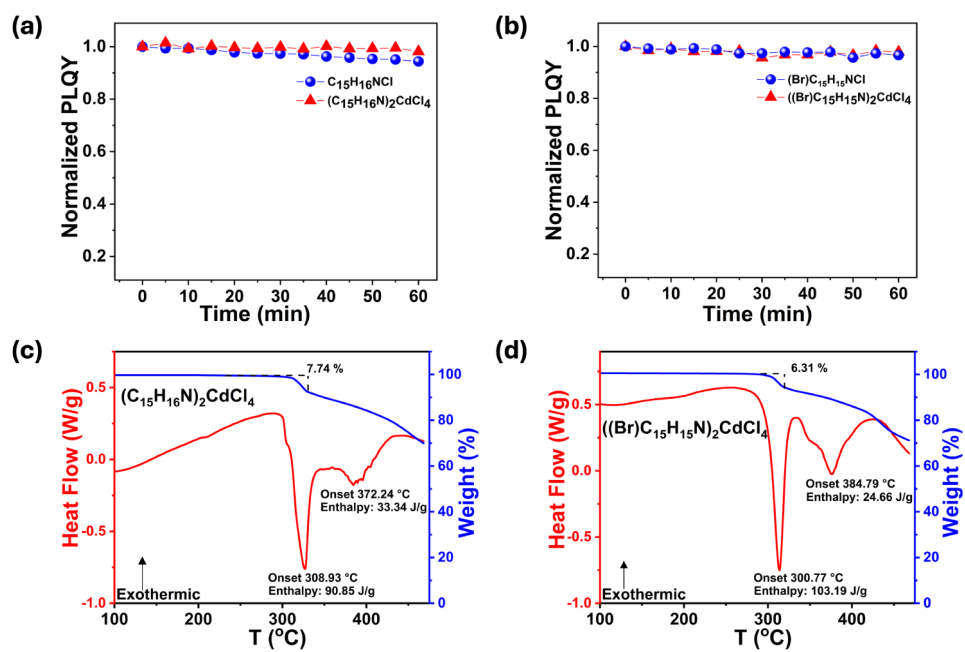

**Figure 5.** Normalized photoluminescence quantum yield (PLQY) under continuous irradiation over 60 min at $PLE_{max}$ for (a) (C$_{15}$H$_{16}$N)$_2$CdCl$_4$ and (b) ((Br)C$_{15}$H$_{16}$N)$_2$CdCl$_4$. Thermogravimetric analysis (TGA) and differential scanning calorimetry (DSC) measurement results for (c) (C$_{15}$H$_{16}$N)$_2$CdCl$_4$ and (d) ((Br)C$_{15}$H$_{15}$N)$_2$CdCl$_4$.

To the best of our knowledge, the PLQY of 26.60% for ((Br)C$_{15}$H$_{15}$N)$_2$CdCl$_4$ is the third highest reported among hybrid cadmium halides; in fact, this value is the second highest if we are to consider only emission from the organic structural component. The lower PLQY of $((Br)C_{15}H_{15}N)_2CdCl_4$ compared to $(C_{15}H_{16}N)_2CdCl_4$ can also be explained through concentration quenching effect; the organic molecular units are spaced closer with comparable intermolecular distances ranging between 5.4 Å and 5.6 Å in $((Br)C_{15}H_{15}N)_2CdCl_4$. These values are in between those of the precursor organics (4.9–5.14 Å) and $(C_{15}H_{16}N)_2CdCl_4$ (7.12 Å–7.42 Å). Consequently, the observed PLQY of 26.60% is more than twice as high as that of the precursor organics but is significantly lower than that of $(C_{15}H_{16}N)_2CdCl_4$. Note that when discussing the lower PLQY of $((Br)C_{15}H_{15}N)_2CdCl_4$, the possible impact of halogen−halogen interactions[77] should also be mentioned. The structural analysis of this compound revealed a shorter distance of 3.82 Å between the cationic layers in the organic bilayer of $((Br)C_{15}H_{15}N)_2CdCl_4$ compared to 4.12 Å in $(C_{15}H_{16}N)_2CdCl_4$. The reduced 3.82 Å distance is within the range of bromine−bromine interaction.[49] This was also confirmed by our DFT calculations—$((Br)C_{15}H_{15}N)_2CdCl_4$ has a noticeably dispersive valence band (see Figure S6) owing to the contribution from Br-$p$ orbitals and overall closer placement of neighboring organic molecules.

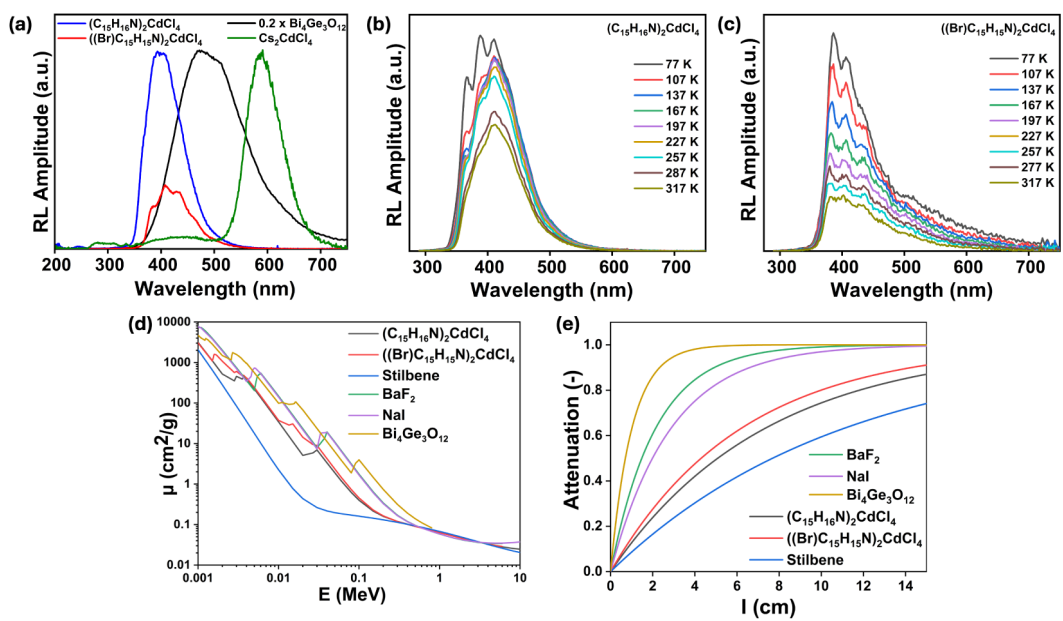

**Figure 6.** (a) Stacked radioluminescence spectra of $(C_{15}H_{16}N)_2CdCl_4$ (in blue), $((Br)C_{15}H_{15}N)_2CdCl_4$ (in red), $Cs_2CdCl_4$ (in green), and $Bi_4Ge_3O_{12}$ (in black). Temperature-dependent radioluminescence spectra for (b) $(C_{15}H_{16}N)_2CdCl_4$ and (c) $((Br)C_{15}H_{15}N)_2CdCl_4$. Attenuation coefficient (d) and attenuation length (e) for $(C_{15}H_{16}N)_2CdCl_4$ (black) and $((Br)C_{15}H_{15}N)_2CdCl_4$ (red) compared to that of some known scintillators including stilbene (blue), BGO (yellow), NaI (purple), and $BaF_2$ (green).

In addition to ACQ leading to lower emission efficiencies, organic light emitters including stilbenes and their derivatives are known to have low photostability due to light-induced changes.[25,26,78,79] To comprehensively characterize the photostability of $(C_{15}H_{16}N)_2CdCl_4$ and $((Br)C_{15}H_{15}N)_2CdCl_4$, PL measurements were performed by irradiating the synthesized materials with their corresponding maximum excitation wavelengths. The precursor organic salts used in this study are noticeably more photostable (Figure 5a-b) than the stilbenes and their derivatives used in other studies;[25,26] the stilbenes used in prior studies showed 60−100% reduction in PLQYs in the same time frame. The photochemical changes (e.g., photoisomerization and photodimerization) of stilbenes are known to quench PL, and for these to occur, the stilbenes must be close to one another (between 3.5 and 4.2 Å).[78] The longer distances (>4.7 Å)[78] such as that observed in $C_{15}H_{16}NCl$ are outside of the range for the photochemical reactions to occur, and therefore, the precursor organics used in the present study are markedly more stable. These organic intermolecular distances are even greater in $(C_{15}H_{16}N)_2CdCl_4$ and $((Br)C_{15}H_{15}N)_2CdCl_4$, and consequently, the PLQYs of both remained almost unchanged after 60 min of continuous UV irradiation (Figure 5a and b). These results suggest that the targeted organic modifications done in this study are effective in improving the photostability of not only $C_{15}H_{16}NCl$ and $(Br)C_{15}H_{15}NCl$, but their respective hybrids as well.

Thermogravimetric analysis (TGA) and differential scanning calorimetry (DSC) measurements (Figure 5c-d) on $(C_{15}H_{16}N)_2CdCl_4$ and $((Br)C_{15}H_{15}N)_2CdCl_4$ suggest that these hybrids have improved thermal stability. Both compounds have onset weight loss temperatures above 300 °C, which were confirmed to be their incongruent melting transitions. The TGA and DSC curves of $(C_{15}H_{16}N)_2CdCl_4$ demonstrate a thermal event at 308.93 °C and an associated weight loss of 7.74% (Figure 5c), which corresponds to the release of $NH_4Cl$. Our hypotheses were supported by the TGA/DSC results of $((Br)C_{15}H_{15}N)_2CdCl_4$, which show a 6.31% (Figure 5d weight loss at around 300 °C, which also corresponds to the loss of a molecule of $NH_4Cl$. These results demonstrate an improved thermal stability of the hybrid compounds compared to not only the precursor organic salts (Figure S12), but also other hybrid metal halides containing stilbene-derived organics show melting or decomposition transitions below 250 °C.[25,26] In $C_{15}H_{16}NCl$, there is a noticeable thermal event at 207.86 °C (Figure S11a). Since heating the sample up to 475 °C shows melting accompanied by decomposition of the samples, we performed additional forward and reverse scans of TGA and DSC of $C_{15}H_{16}NCl$ and $(Br)C_{15}H_{15}NCl$ from 30 to 230 °C. The PXRD data (Figure

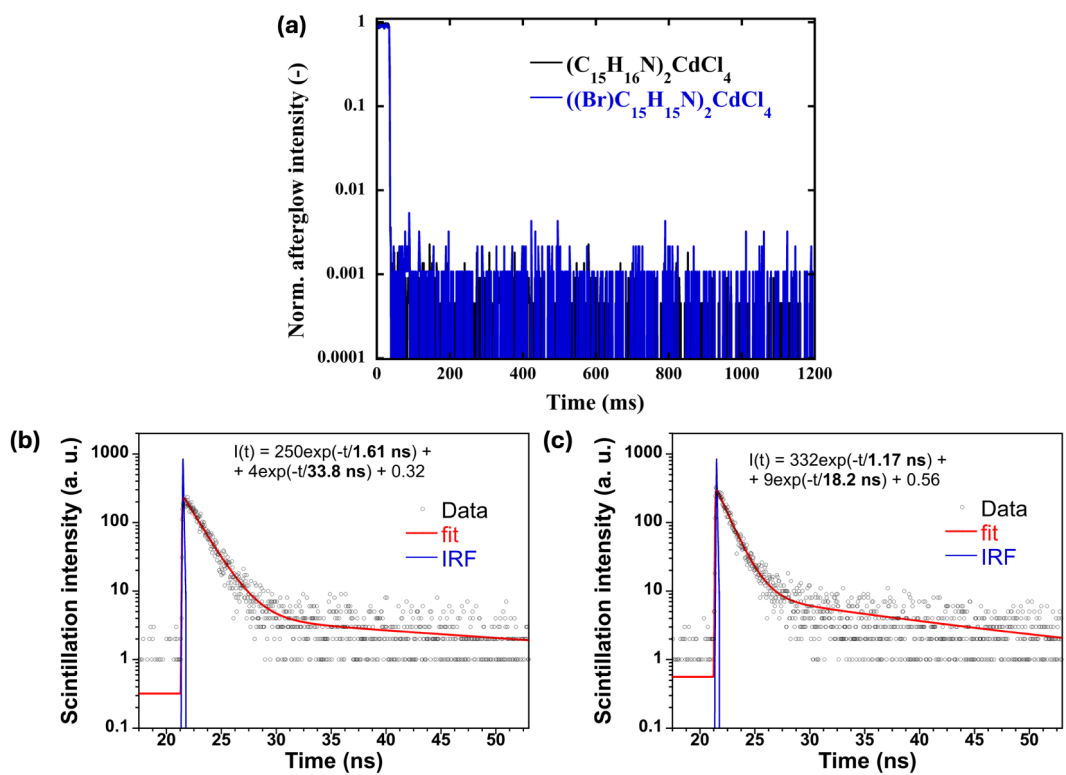

**Figure 7.** (a) Afterglow measurements for $(C_{15}H_{16}N)_2CdCl_4$ and $((Br)C_{15}H_{15}N)_2CdCl_4$. Scintillation decay kinetics for (b) $(C_{15}H_{16}N)_2CdCl_4$ and (c) $((Br)C_{15}H_{15}N)_2CdCl_4$.

S12c,f) obtained for the samples before and after heating cycles show no change, suggesting that any changes in this range are reversible. Importantly, we attribute the improved thermal stability of the hybrids to the presence of 2D layered extended perovskite inorganic sheets in $(C_{15}H_{16}N)_2CdCl_4$ and $((Br)C_{15}H_{15}N)_2CdCl_4$, whereas the prior work was done on 0D molecular crystals.[25,26]

In addition to the improved photostability and thermal stability, $(C_{15}H_{16}N)_2CdCl_4$ and $((Br)C_{15}H_{15}N)_2CdCl_4$ also demonstrate excellent environmental stability. Periodic PXRD measurements taken over a period of 1 year for $(C_{15}H_{16}N)_2CdCl_4$ and over a period of 3 months for $((Br)C_{15}H_{15}N)_2CdCl_4$ show no signs of decomposition or degradation (Figure S13). Altogether, the improved stability and enhanced light emission efficiency, make these materials attractive candidates for potential practical light emission-based applications. Of these, since stilbene-based organics have been previously considered for scintillation applications, the unique combination of advantageous optical and thermal properties and stability of $(C_{15}H_{16}N)_2CdCl_4$ and $((Br)C_{15}H_{15}N)_2CdCl_4$ suggests that they can be excellent candidates for fast radiation detection. For this application, it is important to know if a candidate material can be processed into thin films or composites. We have shown that these novel materials can be processed into films using poly(methyl methacrylate) (PMMA) (Figures S14 and S15). Both films based on $(C_{15}H_{16}N)_2CdCl_4$ and $((Br)C_{15}H_{15}N)_2CdCl_4$ show intense blue emission under 365 nm excitation, and the obtained PL spectra for films are consistent with those for the bulk materials.

### 3.4. Radioluminescence and Scintillation Properties

The radioluminescence (RL) spectra (Figure 6a) closely correlate with the photoluminescence (PL) spectra in terms of spectral shape, suggesting a common emission origin under both UV and X-ray excitation. This interpretation is further supported by the similarity of the decay kinetics observed for UV- and X-ray-excited emissions (Figures 3a-b and 6a−c). The overall RL efficiency, calculated as the integral of the RL spectrum relative to that of a $Bi_4Ge_3O_{12}$ (BGO) reference sample, was determined to be 11% for $(C_{15}H_{16}N)_2CdCl_4$ and 4% for $((Br)C_{15}H_{15}N)_2CdCl_4$. It should be noted that the RL efficiency values for $(C_{15}H_{16}N)_2CdCl_4$ and $((Br)-C_{15}H_{15}N)_2CdCl_4$ are likely underestimated due to the lower collection efficiency of scintillation photons generated deeper within the powder samples, which arises from photon scattering and self-absorption. This effect is particularly pronounced in samples with low X-ray attenuation, i.e., low-density materials composed of lighter elements. To further cement the claim of organic emission, temperature-dependent radioluminescence data were collected for both $(C_{15}H_{16}N)_2CdCl_4$ and $((Br)C_{15}H_{15}N)_2CdCl_4$ (Figure 6b-c). Both $(C_{15}H_{16}N)_2CdCl_4$ and $((Br)C_{15}H_{15}N)_2CdCl_4$ exhibit a clear temperature dependence of the radioluminescence (RL) intensity and spectral shape. In both samples, the overall RL intensity increases as temperature decreases, consistent with suppression of thermally activated nonradiative decay pathways. These likely include intersystem crossing or excited-state ionization. The emission bands become narrower and better resolved at low temperatures, revealing a more pronounced vibronic structure. This can be attributed primarily to reduced band broadening at lower temperatures rather than a change in the underlying electronic transition. Attenuation coefficients (Figure 6d) and attenuation lengths (Figure 6e) of $(C_{15}H_{16}N)_2CdCl_4$ and $((Br)C_{15}H_{15}N)_2CdCl_4$ show their competitiveness with some known scintillators like stilbene, BGO, NaI, and $BaF_2$. Our design concept and synthesis of both materials have demonstrated a substantial increase in the attenuation length in our material when compared to stilbene alone as a scintillator. This is proof of our hypothesis that with further design and engineering, organic–inorganic metal halide-based scintillating materials can rival state-of-the-art pure organic or inorganic-based materials, which suffer from slow response time, sensitivity to temperature, low light yield, and resolution.[80]

The afterglow profiles of both $(C_{15}H_{16}N)_2CdCl_4$ and $((Br)C_{15}H_{15}N)_2CdCl_4$ (Figure 7a) exhibit a rapid decline in signal intensity to below 0.1% of the initial value almost immediately following the termination of excitation. This exceptionally low afterglow, comparable to that of BGO, exceeds the requirements for imaging applications. One of the main challenges faced by today's state-of-the-art scintillators is what is known as ghosting, which is a residual glow or delayed emission that persists after the irradiation source is turned off. The scintillation decay profiles of $(C_{15}H_{16}N)_2CdCl_4$ and $((Br)C_{15}H_{15}N)_2CdCl_4$ were fitted using a biexponential function (Figure 7b-c). In both compounds, the decay is dominated by a fast component with a lifetime closely matching that observed in time-resolved PL measurements, further supporting the assumption of a common emission origin under both UV and X-ray excitation. The slow components could not be accurately fitted due to their low amplitudes. Measurements conducted over an extended time window revealed no evidence of additional, longer-lived decay components, consistent with the results of the afterglow measurements. The light yield ($LY$) of the scintillator can be estimated using a simple formula:

$$LY = \frac{10^6 SQ}{\beta E_g} \text{ph/MeV}$$

where $1/\beta$ is the conversion efficiency, $S$ is the transport efficiency, $Q$ is the quantum efficiency, and $E_g$ is the band gap. To estimate the maximum light yield ($LY_{max}$) of $(C_{15}H_{16}N)_2CdCl_4$ and $((Br)C_{15}H_{15}N)_2CdCl_4$, we can evaluate the formula above assuming ideal transport efficiency ($S = 1$) and the conversion efficiency $1/\beta = 2.5$ typical for wide band gap insulators.[81] For the band gap and quantum efficiency, we can use the estimated band gaps of 3.33 and 3.23 eV from optical absorption spectroscopy (Figure 3c-d) and quantum efficiency 0.5083 and 0.2660 measured for $(C_{15}H_{16}N)_2CdCl_4$ and $Br(C_{15}H_{15}N)_2CdCl_4$, respectively. Based on these values, we get the estimate of maximum achievable light yield of 60,000 ph/MeV and 33,000 ph/MeV for $(C_{15}H_{16}N)_2CdCl_4$ and $Br(C_{15}H_{15}N)_2CdCl_4$, respectively. Based on the fast scintillation decay kinetics and virtually no afterglow of these compounds we can assume direct proportionality between light yield and intensity of radioluminescence ($LY \sim I_{RL}$). Therefore, the new hybrid metal halides obtained in this work can be compared to $Bi_4Ge_3O_{12}$ (BGO) with a typical light yield value around 8,000 ph/MeV. The comparison of obtained values gives a ratio of the theoretical yield of $(C_{15}H_{16}N)_2CdCl_4$ to that of BGO around 7.5. Therefore, these materials have a potential to significantly surpass BGO in terms of light output. On the other hand, based on the results of radioluminescence measurements, the ratio of integral radioluminescence intensity of $(C_{15}H_{16}N)_2CdCl_4$ to BGO is approximately 0.1, corresponding to light yield of 800 ph/MeV. The obtained light yield of 800 ph/MeV for $(C_{15}H_{16}N)_2CdCl_4$ is comparable to the light yield of the fast component of $BaF_2$, a state-of-the-art scintillator which has a light yield of around 1,000 ph/MeV. Therefore, while the obtained results in this work are promising, there is a room for improvement, and with optimization, careful design and molecular engineering, a novel class of even better organic scintillators can be developed with excellent light yield and very low afterglow.

## 4. CONCLUSIONS

In summary, we report on two new compounds, $(C_{15}H_{16}N)_2CdCl_4$ and $((Br)C_{15}H_{15}N)_2CdCl_4$, demonstrating optimized photoemission from organic molecular cations. The materials design approach is generalizable and is based on combining nonemissive inorganic component, $[CdCl_4]^{2-}$, together with photoemissive conjugated organic molecules. The new hybrid halides have 2D layered Ruddlesden–Popper-type perovskite structures featuring inorganic $[CdCl_4]^{2-}_\infty$ layers that provide a rigid extended framework to which organic cations are anchored in fixed distances. The elongated distances between the organic molecules in $(C_{15}H_{16}N)_2CdCl_4$ and $((Br)C_{15}H_{15}N)_2CdCl_4$, as compared to their respective precursor organic salts, are very important in preventing ACQ, and result in excellent PLQYs of 50.83% and 26.60% for $(C_{15}H_{16}N)_2CdCl_4$ and $((Br)C_{15}H_{15}N)_2CdCl_4$, respectively. The observed multiband blue emission in $(C_{15}H_{16}N)_2CdCl_4$ and $((Br)C_{15}H_{15}N)_2CdCl_4$ is similar to that of their corresponding precursor salts $C_{15}H_{16}NCl$ and $(Br)C_{15}H_{15}NCl$ but with an important distinction. The phosphorescence observed in the 500–600 nm spectral region observed for the organic salts are suppressed in the hybrids due to the increased rigidity of the hybrid structure owing to the presence of heavy inorganic metal halide $[CdCl_4]^{2-}_\infty$ layers. Yet another benefit of the increased organic–organic intermolecular distances and structural rigidity, $(C_{15}H_{16}N)_2CdCl_4$ and $((Br)C_{15}H_{15}N)_2CdCl_4$ exhibit improved photostability and thermal stability not only as compared to their respective precursor organic salts but also other 0D hybrid metal halides employing stilbene-based organic cations.[25,26]

The unique combination of efficient and fast light emission from the organic structural component and improved stability allows consideration of $(C_{15}H_{16}N)_2CdCl_4$ and $((Br)-C_{15}H_{15}N)_2CdCl_4$ for practical applications. Among them, there has been an increasing demand for advanced scintillator technologies in recent years.[82] Of these, materials exhibiting very fast scintillation decay kinetics are particularly relevant for applications such as time-of-flight positron emission tomography (TOF-PET),[83] computed tomography employing photon-counting detectors,[84] and high-energy physics experiments.[85] Although there is a growing interest from the scientific community, the potential of hybrid organic–inorganic halides in these domains remains underexplored. This work demonstrates that $(C_{15}H_{16}N)_2CdCl_4$ and $((Br)-C_{15}H_{15}N)_2CdCl_4$ can integrate high attenuation of high-energy photons, enabled by the presence of inorganic metal halide sheets with the fast ns scintillation decay kinetics characteristic of organic molecular systems. The estimated light yield of 800 ph/MeV for $(C_{15}H_{16}N)_2CdCl_4$ is very encouraging, and further improvements should be possible with fine-tuning of chemical composition and crystal structure. For instance, future work could target further optimization of distances between organic emitters through the modification of the organic emitters and/or changes to the inorganic layers (e.g., halide alloying). Yet another focus area for future research, if the charge transfer efficiency between the inorganic sheets and organic emitters can be sufficiently improved, the resultant materials could significantly outperform the current benchmark $BaF_2$[85−87] in terms of coincidence time resolution, owing to the combined advantages of higher light output and superior spectral matching with silicon photomultipliers (SiPMs). Altogether, the reported results in this work suggest that hybrid organic–inorganic metal halides can be compositionally and structurally engineered to have highly efficient photoemission originating from the organic components and that such materials are excellent candidates for fast radiation detection applications.

## ■ ASSOCIATED CONTENT

### *ⓈSupporting Information

The Supporting Information is available free of charge at https://pubs.acs.org/doi/10.1021/jacs.5c20638.

> Experimental details for the synthesis of organic precursors, together with NMR analysis, supplementary crystallographic data, PXRD patterns, PL spectra, and computational results (PDF)

### Accession Codes

Deposition Numbers 2463650 and 2463654−2463655 contain the supplementary crystallographic data for this paper. These data can be obtained free of charge via the joint Cambridge Crystallographic Data Centre (CCDC) and Fachinformationszentrum Karlsruhe Access Structures service.

## ■ AUTHOR INFORMATION

### Corresponding Author

**Bayram Saparov** − *Department of Chemistry & Biochemistry, The University of Oklahoma, Norman, Oklahoma 73019, United States;* orcid.org/0000-0003-0190-9585; Email: saparov@ou.edu

### Authors

**Muhammad S. Muhammad** − *Department of Chemistry & Biochemistry, The University of Oklahoma, Norman, Oklahoma 73019, United States;* orcid.org/0009-0008-4685-9087

**Dilruba A. Popy** − *Department of Chemistry & Biochemistry, The University of Oklahoma, Norman, Oklahoma 73019, United States;* orcid.org/0000-0001-5017-3274

**Hamza Shoukat** − *Department of Chemistry & Biochemistry, The University of Oklahoma, Norman, Oklahoma 73019, United States*

**John M. Lane** − *Dave C. Swalm School of Chemical Engineering and Center for Advanced Vehicular Systems, Mississippi State University, Starkville, Mississippi 39762, United States*

**Neeraj Rai** − *Dave C. Swalm School of Chemical Engineering and Center for Advanced Vehicular Systems, Mississippi State University, Starkville, Mississippi 39762, United States;* orcid.org/0000-0002-0058-9623

**Vojtěch Vaněček** − *Institute of Physics, Academy of Science of the Czech Republic, Praha 16200, Czech Republic;* orcid.org/0000-0001-9730-9570

**Zdeněk Remeš** − *Institute of Physics, Academy of Science of the Czech Republic, Praha 16200, Czech Republic;* orcid.org/0000-0002-3512-9256

**Romana Kučerková** − *Institute of Physics, Academy of Science of the Czech Republic, Praha 16200, Czech Republic;* orcid.org/0000-0001-9441-0681

**Vladimir Babin** − *Institute of Physics, Academy of Science of the Czech Republic, Praha 16200, Czech Republic;* orcid.org/0000-0003-3072-2242

**Chenjia Mi** − *Department of Chemistry & Biochemistry, The University of Oklahoma, Norman, Oklahoma 73019, United States;* orcid.org/0000-0003-2169-864X

**Yitong Dong** − *Department of Chemistry & Biochemistry, The University of Oklahoma, Norman, Oklahoma 73019, United States;* orcid.org/0000-0002-7069-3725

**Mark D. Smith** − *Department of Chemistry and Biochemistry, University of South Carolina, Columbia, South Carolina 29208, United States*

**Novruz G. Akhmedov** − *Department of Chemistry & Biochemistry, The University of Oklahoma, Norman, Oklahoma 73019, United States*

**Daniel T. Glatzhofer** − *Department of Chemistry & Biochemistry, The University of Oklahoma, Norman, Oklahoma 73019, United States*

Complete contact information is available at:
https://pubs.acs.org/10.1021/jacs.5c20638

### Author Contributions

The manuscript was written through the contributions of all authors. All authors have given approval of the final version of the manuscript.

### Notes

The authors declare no competing financial interest.

## ■ ACKNOWLEDGMENTS

This work was supported by the U.S. Department of Energy, Office of Science, Office of Basic Energy Sciences, under Award No. DE-SC0021158. The work is additionally supported by OP JAC financed by ESIF and MEYS (Project LASCIMAT─CZ.02.01.01/00/23_020/0008525). The fluorescence lifetime measurements were supported by the U.S. Department of Energy, Office of Science, Office of Basic Energy Sciences, under Award No. DE-SC0024441. This research used resources of the National Energy Research, Scientific Computing Center, a DOE Office of Science User Facility supported by the Office of Science of the U.S. Department of Energy under Contract No. DE-AC02-05CH11231 using NERSC award NERSC DDR-ERCAP0034551.

## ■ REFERENCES

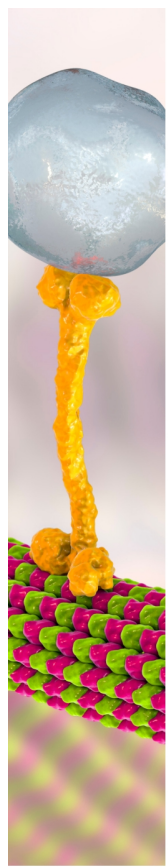